\documentclass{aastex61}

\submitjournal{ApJ}

\shorttitle{Muon bundles from strangelets}
\shortauthors{Kankiewicz et al.}

\begin{document}

\title{Muon bundles as a sign of strangelets from the Universe}

\correspondingauthor{M. Rybczy\'{n}ski}
\email{pawel.kankiewicz@ujk.edu.pl, maciej.rybczynski@ujk.edu.pl, zbigniew.wlodarczyk@ujk.edu.pl, grzegorz.wilk@ncbj.gov.pl}

\author{P. Kankiewicz}
\affiliation{Institute of Physics, Jan Kochanowski University, 25-406 Kielce, Poland}

\author{M. Rybczy\'{n}ski}
\affiliation{Institute of Physics, Jan Kochanowski University, 25-406 Kielce, Poland}

\author{Z. W\l odarczyk}
\affiliation{Institute of Physics, Jan Kochanowski University, 25-406 Kielce, Poland}

\author{G. Wilk}
\affiliation{National Centre for Nuclear Research, Department of Fundamental Research, 00-681 Warsaw, Poland}

\begin{abstract}

Recently the CERN ALICE experiment, in its dedicated cosmic ray run, observed muon bundles of very high multiplicities, thereby confirming similar findings from the LEP era at CERN (in the CosmoLEP project).
Originally it was argued  that they apparently stem from the primary cosmic rays with a heavy masses.
We propose an alternative possibility arguing that muonic bundles of highest multiplicity are
produced by strangelets, hypothetical stable lumps of strange quark matter infiltrating our Universe.  We also address the possibility of additionally deducing their directionality which could be of astrophysical interest. Significant evidence for anisotropy of arrival directions of the observed high multiplicity muonic bundles is found. Estimated directionality suggests their possible extragalactic provenance.

\end{abstract}


\section{Introduction} \label{sec:intro}

Cosmic ray physics is our unique source of information on events in the energy range which will never be accessible in Earth-bound experiments~\cite{DR,L-SS}. This is why one of the most important aspects of their investigation is the understanding of the primary cosmic ray (CR) flux and its composition. In this respect the recent measurement performed by the ALICE experiment at CERN LHC in its dedicated cosmic ray run~\cite{ALICE:2015wfa,ALICE:2015wfa_suppl} is of great importance. A number of events with muon bundles of high multiplicity was registered in the so called Extensive Air Showers (EAS) produced by cosmic ray interactions in the upper atmosphere~\cite{ALICE:2015wfa}. A special emphasis has been given to the study of high multiplicity events containing more than $100$ reconstructed muons. Similar events have already been studied in the previous LEP experiments at CERN (in the so called CosmoLEP program) such as ALEPH~\cite{Avati:2000mn}, DELPHI~\cite{Abdallah:2007fk} and L3~\cite{Achard:2004ws}, and in the underground Baksan experiment~\cite{Baksan} (we do not mention here all experiments with smaller bundles of muons observed). Data obtained in all these studies are crucial in establishing the most probable composition of CR. The recent ALICE experiment was special in this respect because it also provides, for the first time, data which can be used to estimate the direction of arrival of the observed CR producing observed muonic bundles~\cite{ALICE:2015wfa}.

The common feature shared by data from all these experiments is that the measured muon multiplicity distribution can be reproduced only for low and intermediate multiplicities of the produced muons. This is done using Monte Carlo simulations assuming some suitable combination of the protonic and iron components in the incoming CR. However, these simulations encounter difficulties in description of the frequency distribution of the highest multiplicity events (containing more than 100 muons in a bundle). The situation is best summarized by noting that the observation of  high multiplicity muon bundles in CR events at CosmoLEP is apparently the only result that does not agree with the Standard Model. The measured multiplicity distribution of muons produced by high energy CR appear to be very sensitive both to the composition of the CR flux entering our atmosphere and to the assumptions concerning the dominant hadronic production mechanisms in air shower development~\cite{ALICE:2015wfa}. In fact, in \cite{ALICE:2015wfa} it is argued that bundles of very high multiplicity of muons stem from primary cosmic rays with energies exceeding  $10^{16}$ eV and that their frequency can be described by Fe component of primary cosmic rays in this energy range. This is the first time that the rate of high multiplicity muon bundles has been reproduced using a conventional hadronic model (adopted CORSIKA event generator version 7350 and incorporating QGSJET II-04 model in which pion exchange is assumed to dominate forward neutral hadron production) for the description of extensive air shower. However, when comparing their Figures 5 and 6  one notices that, whereas the first  presents the measured muon multiplicity distribution of the whole sample of data, the second presents a CORSIKA fits to measured data limited only to the intermediate muon multiplicities, $N_{\mu} \sim 70$. One can see that  data for low multiplicity are located in the middle between the model predictions corresponding to the pure p and pure Fe compositions. If one tries to extrapolate ALICE fits to the end of the measured distribution it turns out that, even for the pure iron  fit, there is a noticeably disagreement with the measured points, of about one order of magnitude, depending on the extrapolation method used.

In this paper we propose an alternative possibility arguing that muonic bundles of highest multiplicity are  produced by strangelets. We shall concentrate mainly on the first, high multiplicity events. To describe them we propose seriously to consider, for a moment, the possibility of the existence in the flux of incoming CR a component with very high atomic number, of the order of $A \sim 10^3$. In fact we propose to return to our old idea that muon bundles of extremely high multiplicities could be produced by \emph{strangelets}, hypothetical stable lumps of strange quark matter (SQM) infiltrating our Universe~\cite{Ryb}. Strangelets with such masses, much larger than the masses of ordinary nuclei, could easily (without invoking any peculiar form of interactions) produce extremely large groups of muons in collisions with the atmosphere. In the next Section we list and discuss briefly the searches for strangelets performed so far and provide the corresponding limits of their appearance. In Section \ref{sec:HighN} we briefly recall our arguments as to why strangelets can penetrate quite deeply into the Earth's atmosphere, provide our results for the high multiplicity muon bundles they can produce and compare them with the recent ALICE data. By way of illustration we shall present there our old results obtained for data registered in the old CosmoLEP project at CERN. We demonstrate that already a relatively minute (of the order of $10^{-5}$ of the total primary flux) admixture of the SQM in the CR of the same total energy is enough for this purpose and that it is consistent with all recent observations. Section \ref{sec:Anisotropy} is devoted to estimations of the possible direction from which strangelets producing our muon bundles could arrive. Section \ref{sec:conc} contains our conclusions.

\section{Searches for Strange Quark Matter in cosmic rays}
\label{sec:sec-1}

We start with a short reminder of strange quark matter and strangelets as seen from the present perspective. The idea of SQM originated some time ago in \cite{HT,Witten,FJ,Alcock:1985vc}, but it remains alive, cf. \cite{Madsen-Q,S-MGA,S-FW,S-RX,S-RK,HT-NPCS1,HT-NPCS2}\footnote{In fact it also became a source of new inspirations, as witnessed, for example, by \cite{ST-ACTA,ST-CWS,RN,SQM-DM,SL-TChin}.}. In short, the basic, commonly accepted view is that that SQM (understood as a combination of roughly an equal number of up, down and strange quarks) might be the true ground state of quantum chromodynamics (QCD). It is therefore reasonable to expect that it exists in some form in the Universe and can be detected. This supposition resulted in a number of searches for strange stars and quark stars, in which such a form of matter would be dominant and which could therefore be a possible source of strangelets penetrating outer space \cite{SS-AFO,QS-IJMPD,QS-Science,QS-Drake,QS-PRL,SL-EF,QS-MNRAS,QS-AA}.

Of special interest to us are strangelets, lumps of SQM with baryon number exceeding some critical value, $A > A_{crit}\sim 300 - 400$~\cite{Alcock:1985vc}. In this region of $A$ they are absolutely stable against neutron emission, below this limit they decay rapidly by evaporating neutrons. As shown in \cite{Wilk:1996me,Wilk:1996jpg} they can penetrate deep into the Earth's atmosphere, notwithstanding their very large initial masses and sizes.  It is therefore fully sensible to search for them in the CR experiments \cite{Wilk:1996je,Ryb:2002mww,Ryb:2006mww,Ryb:2004rr}. Not going into details, we argued in the following way. The geometrical radii of strangelets are estimated to be comparable to the radii of ordinary nuclei~\cite{Wilk:1996me}, which means that their geometrical cross sections are similar to the normal nuclear ones. The first thing that comes to mind is that this is in contradiction with their strong penetrability which we need for our purpose. However, the fact that strangelets are not normal nuclei but rather a kind of quark bag, allows us to propose the following possible scenario. Namely, we can envisage that strangelets reaching deeply into the atmosphere are formed in many successive interactions with air nuclei by the initially very heavy lumps of SQM entering the atmosphere and decreasing in mass due to collisions with air nuclei (until their $A$ reaches the critical value $A_{crit}$~\cite{Wilk:1996me} and they decay rapidly). It turns out that such a scenario is fully consistent with all present experiments~\cite{Wilk:1996me}. In this scenario the interaction of strangelet with a target nucleus involves all the quarks of the target located in the geometrical intersection of the colliding air nucleus and the strangelet.

\begin{figure}[t]
\begin{center}
\includegraphics[width=10cm]{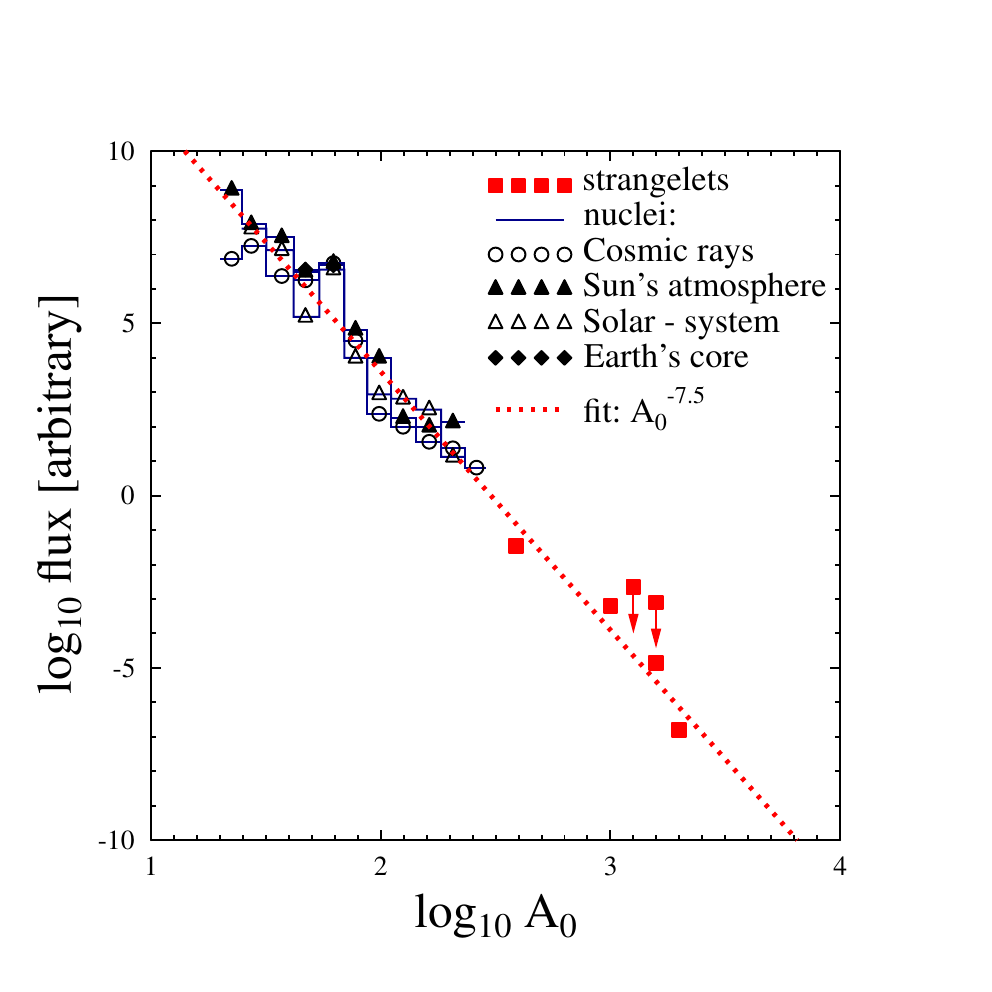}
\caption{\label{fig:comp} Comparison of the estimated mass spectrum $N(A_0)$ of strangelets with the known abundances of elements in the Universe \cite{Z}. Consecutive steps in the histogram denote the following nuclei (or groups of nuclei): Ne, (Mg, Si), S, (K, Ca), Fe, (Cu, Zn), (Kr, Sr, Zr), (Te, Xe, Ba), (rare~earths), (Os, Ir, Pt, Pb). The flux of strangelets has been chosen to accommodate as much as possible all signals of strangelets \cite{Ryb:2002mww,Ryb:2004rr}.}
\end{center}
\end{figure}
\begin{figure}[h]
\begin{center}
\includegraphics[width=10cm]{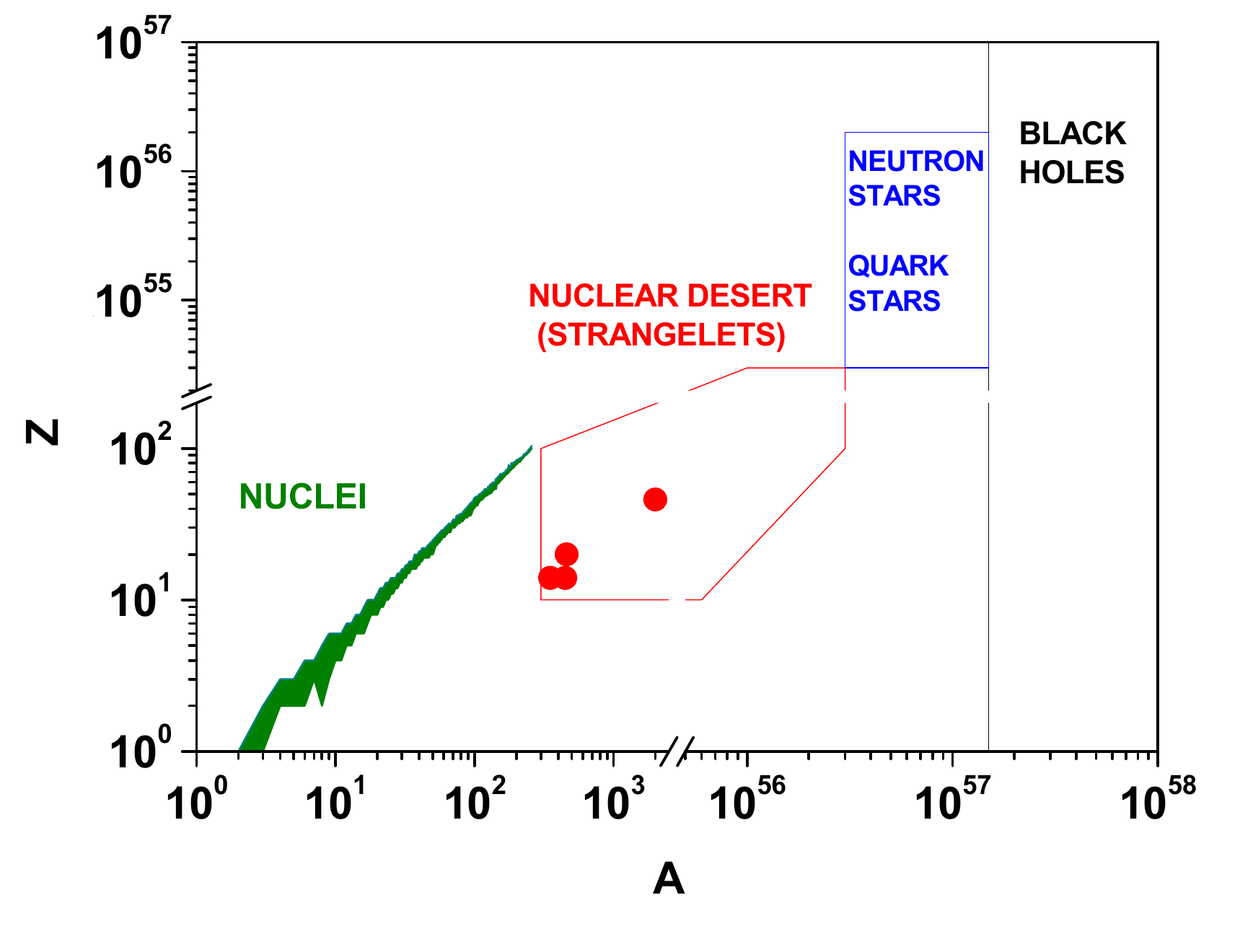}
\caption{Chart of the nuclides showing all known forms of stable matter. Unpopulated nuclear desert, between heaviest nuclides and neutron (or quark) stars, may be filled with strangelets.
}
\label{fig:chart}
\end{center}
\end{figure}
\begin{figure}[h]
\begin{center}
\includegraphics[width=10cm]{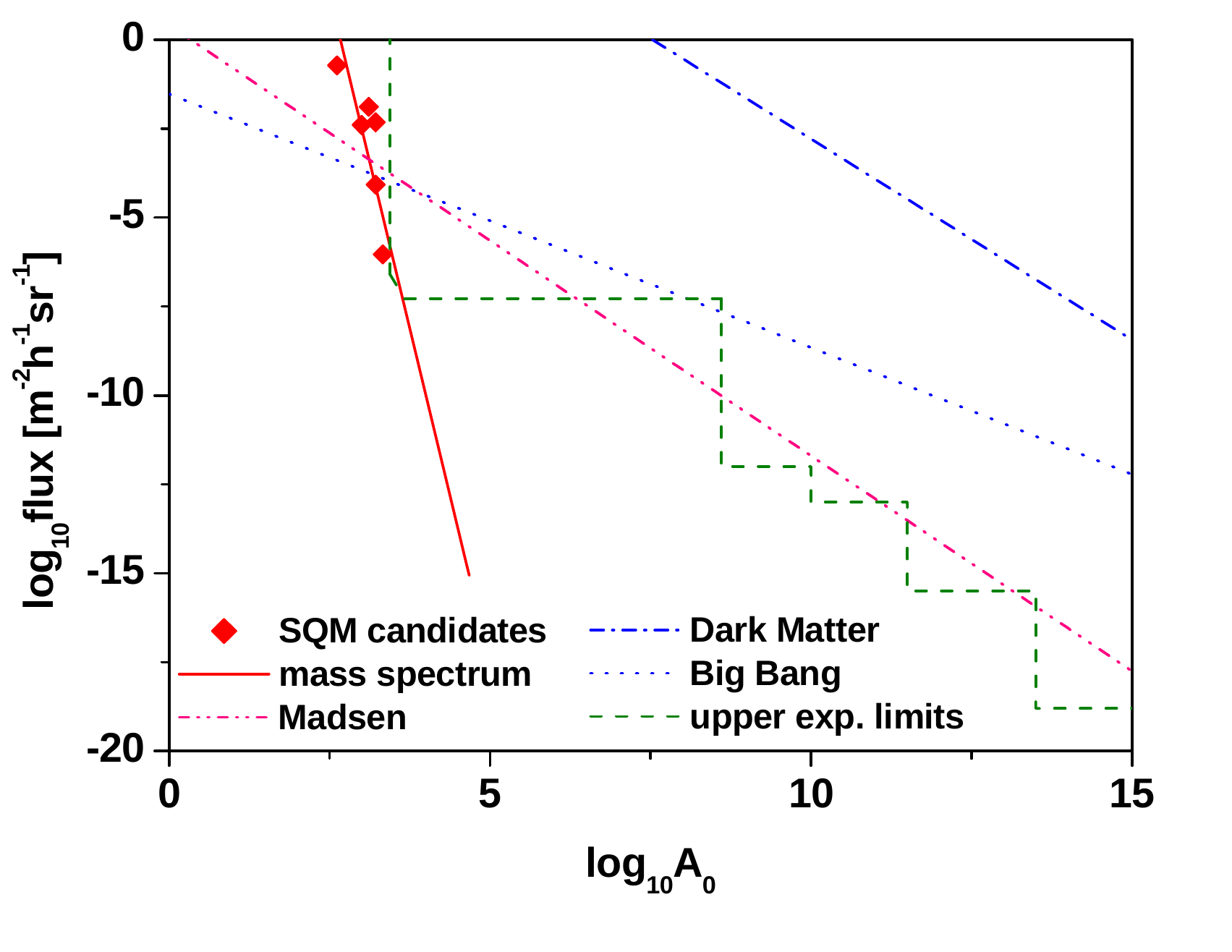}
\caption{The expected flux (our results \cite{Ryb:2002mww}) of strangelets compared with the upper
experimental limits compiled by Price~\cite{Price:1988ge} and SLIM~\cite{Sahnoun:2008mr}. The expected flux from strangelet production in strange star collisions (Madsen spectrum~\cite{Madsen}) and the predicted astrophysical limits \cite{Price:1988ge} are shown.}
\label{fig:chart-1}
\end{center}
\end{figure}

This scenario of large strangelets arriving from outer space and colliding with the Earth's atmosphere is the only one that leads to the observed pattern of production of high multiplicity muonic bundles. Nevertheless, it should be noted that there exists the reverse scenario in which a rather small impinging strangelet picks up mass from the atmospheric atoms when traversing the atmosphere \cite{Raha-1,Raha-2}. However, in this case it would be impossible to get bunches of muons of the observed multiplicity\footnote{It should also be mentioned that there are also a number of other approaches, which we shall not discuss here, like, for example, \cite{Madsen,SL-P,Monreal,PH}. A unified theoretical description of objects ranging from strangelets to strange stars has been recently proposed in \cite{S-to-SS}.}.

There are several reports suggesting the existence of direct candidates for SQM~\cite{Saito:1990ju} (characterized mainly by their very small ratios of $Z/A$). All of them have mass numbers $A$ near or slightly exceeding $A_{crit}$. Analysis of these candidates for SQM shows~\cite{Wilk:1996me} that the abundance of strangelets in the primary cosmic ray flux is $F_{S}\left(A=A_{crit} = 320\right)/F_{tot}\sim 2.4\cdot 10^{-5}$ at the same energy per particle. For the normal flux of primary cosmic rays~\cite{Shibata:1995xz} the expected flux of strangelets is then equal to $F_{S}=7\cdot 10^{-6}~{\rm m}^{-2}{\rm h}^{-1}{\rm sr}^{-1}$ for an energy above $10$~GeV per initial strangelet\footnote{Note that there is a well known equality between the energy density of cosmic rays, $\rho_{CR}$, the magnetic fields, the motion of gas clouds and  starlight. The estimated value of $\rho_{CR} \sim 0.8$ eV/cm$^3$, is comparable with the energy density of the cosmic microwave background, $\rho_{CMB} \sim 0.3$ eV/cm$^3$ \cite{TWAW}. On the other hand, the energy density of strangelets, $\rho_{SQM} \sim 2\cdot 10^{-6}$ eV/cm$^3$, is apparently of the same order as that of the fluctuations in the CMB. }. These estimations of SQM flux do not contradict the results obtained recently by the SLIM Collaboration~\cite{Sahnoun:2008mr}.

To summarize:
\begin{itemize}
\item The experimental data mentioned above lead to a flux of strangelets which follows the $A^{-7.5}$ behaviour, which in turn coincides with the behaviour of the abundance of normal nuclei in the Universe, see Figure~\ref{fig:comp}.
\item Strange quark matter fills a vast gap in the distribution of all known forms of stable matter, cf. chart of nuclides presented in~\cite{Crawford:1994cn} and shown in Figure~\ref{fig:chart}. It places itself exactly between the heaviest atomic elements and neutron stars.
\item The estimated flux of strangelets is also consistent with the astrophysical limits and with the upper limits given experimentally~\cite{Price:1988ge}, see Figure~\ref{fig:chart-1}.

\end{itemize}

\section{High multiplicity muon bundles from SQM}
\label{sec:HighN}

We start with a short reminder of our proposed possible scenario of propagation of strangelets through the atmosphere \cite{Wilk:1996me,Wilk:1996jpg}. The apparent contradiction between the large initial size of the incoming strangelets  and their required strong penetrability in the atmosphere can be resolved by assuming that strangelets reaching deeply into the atmosphere are formed from the original large strangelet which loses mass in many successive interactions with air nuclei when penetrating the atmosphere. To provide numerical estimate we limit ourselves to the two most extreme pictures of the collision of a strangelet of mass number $A$ with an air nucleus target of mass number $A_t$:
\begin{figure}[t]
\begin{center}
\includegraphics[width=15cm]{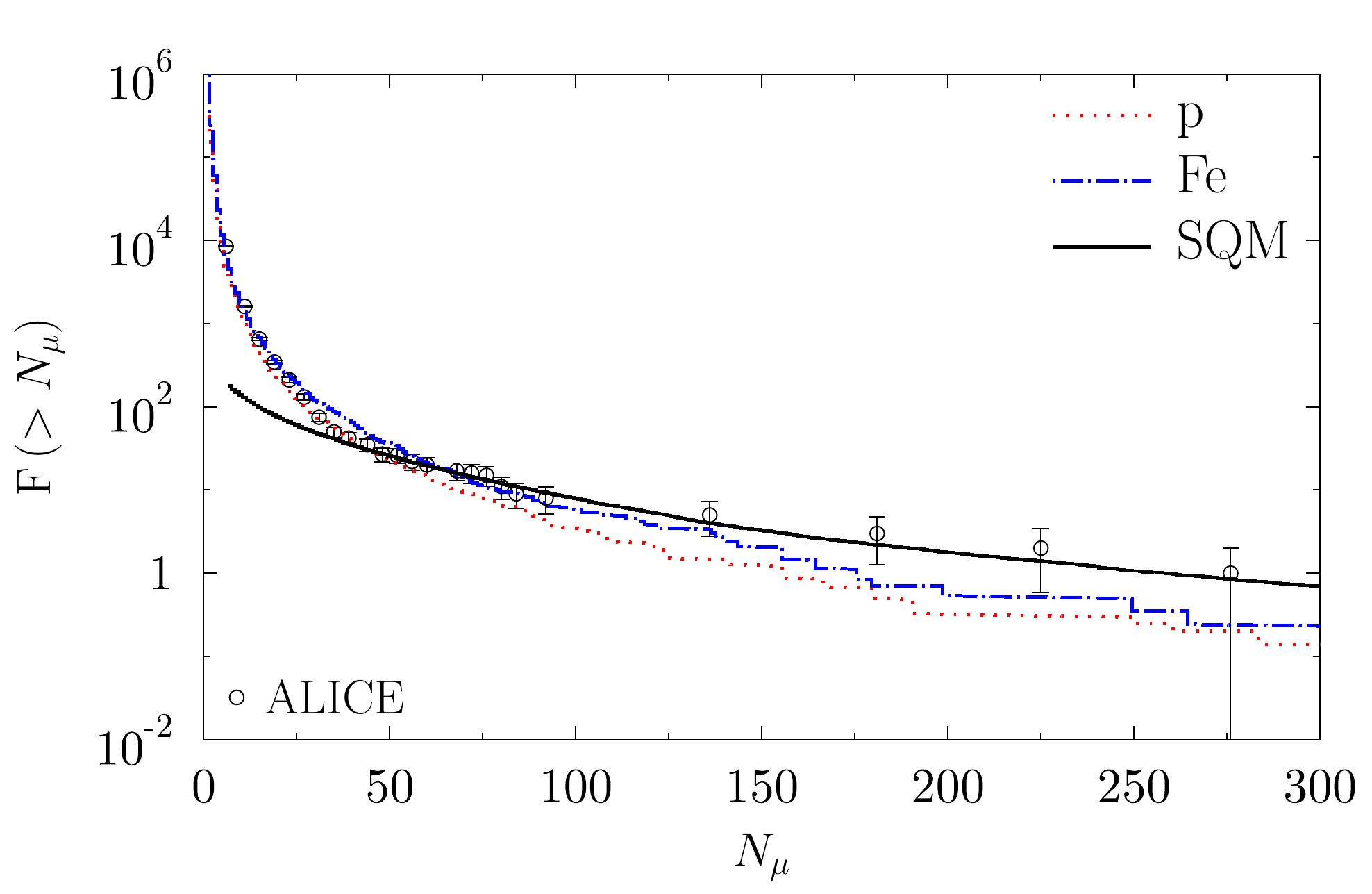}
\caption{Integral multiplicity distribution of muons for the ALICE data~\cite{ALICE:2015wfa} (circles). Monte Carlo simulations for primary protons (dotted line), iron nuclei (dashed dot line) and for the primary strangelets
with mass $A$ taken from the $A^{-7.5}$  distribution (full line) with abundance (at $A=A_{crit}$) $2 \cdot 10^{-5}$ of the total primary flux.}
\label{fig:hmm}
\end{center}
\end{figure}
\begin{figure}[t]
\begin{center}
\includegraphics[width=15cm]{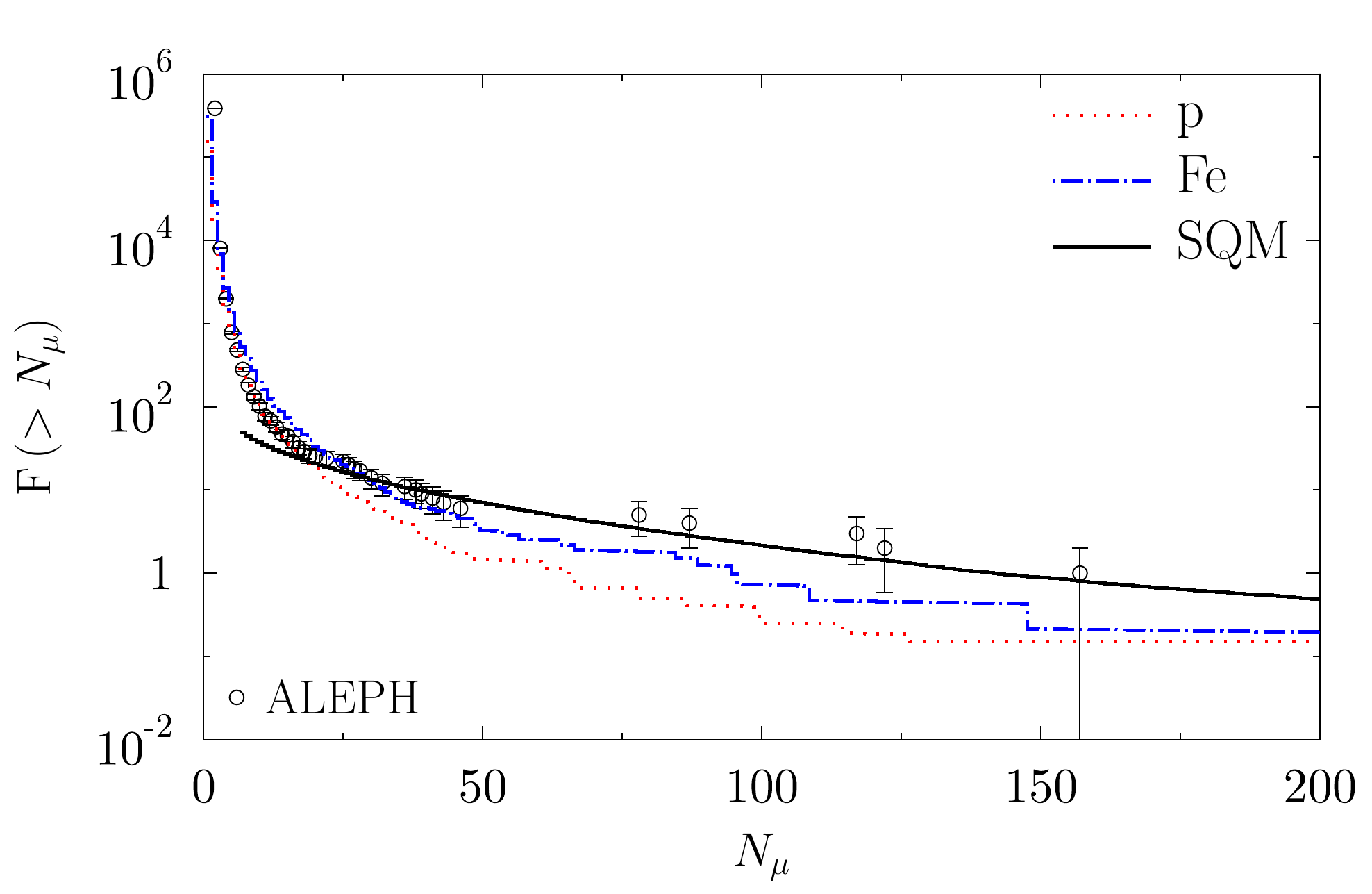}
\caption{Integral multiplicity distribution of muons for the CosmoLEP data (circles) \cite{Avati:2000mn} compared with contributions from primary protons (dotted line), iron nuclei (dashed dot line) and primary strangelets (full line) with mass $A$ taken from the $A^{-7.5}$ distribution and with abundance (at $A=A_{crit}$) $2 \cdot 10^{-5}$ of the total primary flux.}
\label{fig:hmm-1}
\end{center}
\end{figure}
\begin{itemize}
\item[$(i)$] All quarks of $A_t$ which are located in the geometrical intersection of the two colliding projectiles are involved and one assumes that each quark from the target interacts with only one quark from the strangelet.
\item[$(ii)$] All quarks from both nuclei which are in their geometrical intersection region participate in the collision.
\end{itemize}

In the first case, during the interaction up to ($3A_t$) quarks from the strangelet could be used up and its mass could drop to a value of $A - A_t$, at most. The total penetration depth of the strangelet can be estimated to be in this case equal to
\begin{equation}
\Lambda \simeq \frac{1}{3}\lambda_{NA_t}\left( \frac{A_0}{A_t}\right)^{\frac{1}{3}}\left( 1 - \frac{A_{crit}}{A_0}\right)^2 \left( 4 - \frac{A_{crit}}{A_0} \right) \simeq \frac{4}{3} \lambda_{NA_t}\left( \frac{A_0}{A_t}\right)^{\frac{1}{3}} \label{eq:Scen1}
\end{equation}
(here $\lambda_{NA_t}$ denotes the mean free path for $N-A_t$ collisions). One accommodates here both the most probable "normal" mean free paths for successive interactions and the final large penetration depth. This scenario is fully consistent with all present and proposed experiments and could be additionally checked only by measuring the  products of the intermediate collisions, which so far is impossible.

The second case is an analogue of the so called tube model used occasionally in nuclear collisions (cf., for example, \cite{FT}). After the collision the atomic mass of the strangelet diminishes to a value equal to $A -  A^{1/3}\cdot A_t^{2/3}$.  While this is a rather extreme variant, it is still useful in providing  an estimate of the maximum possible destruction of the quarks in the strangelet.

In this study we have used suitably modified SHOWERSIM~\cite{SHOWERSIM84} modular software. We performed Monte Carlo simulations for primary nuclei (protons and  iron nuclei) and for primary strangelets with mass $A$ taken from the   $A^{-7.5}$ distribution. We have generated events initiated by the interaction of primaries with zenith angles in the interval $0 \leq \theta \leq 50^{\circ}$  and with the energies $E\geq 10^{14}$ eV, sampled from the power energy spectrum $F(E) \propto E^{-2.8}$. The total flux of cosmic rays was assumed to be $F\left(1~{\rm TeV}\right)=0.3~{\rm m}^{-2}{\rm s}^{-1}{\rm sr}^{-1}{\rm TeV}^{-1}$. In the analysis were considered muons in the nuclear cascade reaching the ALICE  surface location ($450$ m above sea  level). Our simulations did not explicitly propagate muons in the rock above the ALICE detector (located $52$ meters underground with $28$ meters of overburden  rock, what corresponds to a momentum cut-off of $16$ GeV for the vertical incidence of muons). Instead, the trajectories of muons with energies $E_{\mu} \geq 16$ GeV/$\cos\left(\theta\right)$ arriving at the surface were extrapolated as straight lines  to the depth of the detector. Each shower was randomly scattered over an area $200\times 200$ m$^2$ centered on the ALICE detector. All muons passing the energy cut and crossing an effective detector area of $17$ m$^2$  were considered as detected. The results of our simulations (equivalent to 30.8 days live time to permit direct comparison with the date without the need to apply an arbitrary normalization factor) are shown in Figure~\ref{fig:hmm}. Note that whereas the lower and medium multiplicities can be reproduced by the ordinary nuclei, the extremely large groups of muons can be described only by allowing some admixture of the SQM of the same total energy (a relatively minute one, of the order of the $10^{-5}$  of the total primary flux). Figure~\ref{fig:hmm-1} shows, for comparison, similar results obtained for the CosmoLEP data (recorded by the ALEPH detector which was located at the deepest LEP point, $140$ m underground, corresponding to a energy cut-off of $70$ GeV for vertical incidence of muons) \cite{Avati:2000mn}. Integral multiplicity distributions of muons with energies $E_{\mu} \geq 70$~GeV/$\cos\left(\theta\right)$, crossing an detector area of 16~$m^{2}$ and correspond to 20.2 days effective run time in which showers within the angular range $0\leq \theta\leq 60^{\circ}$ have been collected.

\section{Anisotropy of arrival directions of strangelets}
\label{sec:Anisotropy}

The ALICE data also turns out to be very valuable for another reason. Namely, they show the angular distribution of the muon events in the spherical reference frame with zenith angle ($\theta$) and azimuth angle ($\Phi$) (see \cite{ALICE:2015wfa} and \cite{LHCdesign:2004} for the detailed orientation of the system of coordinates used). The $Y$ axis is directed to the Zenith, and the orientation of the $X$ axis is the same as most installations at ALICE Point $2$, especially the buildings located perpendicularly to the SXL2 and SX2 halls in the surface zone. We can therefore determine the angle between the $X$ axis and the geographical direction to the north: $\alpha_{0}=56.12^{\circ}$.

\begin{table}[h]
\caption{Celestial coordinates of five high-multiplicity muon events. Column description: Event: the label of the event, $N_{\mu}$: the number of registered atmospheric muons, $\Phi$, $\theta$: azimuth and zenith angles (source: \cite{ALICE:2015wfa} and supplemental figures \cite{ALICE:2015wfa_suppl}), Date (JD): the Julian date of the event calculated from the timestamp of each event, $Az$: the Azimuth measured from the north, $h$: the elevation above the horizon, $\alpha_{2000}$ and $\delta_{2000}$: equatorial coordinates related to the $J2000.0$ epoch.}\label{Table1}
\begin{center}
\begin{tabular}{|c|c|c|c|c|c|c|r|r|}
\hline
Event & $N_{\mu}$ & $\Phi$ [$^{\circ}$] & $\theta$ [$^{\circ}$] & Date (JD) & $Az$ [$^{\circ}$] & $h$ [$^{\circ}$] & \multicolumn{1}{c|}{$\alpha_{2000}$} & \multicolumn{1}{c|}{$\delta_{2000}$} \\
\hline
1 & 181 & 212.4 & 40.4 & 2455247.53666 & 268.5 & 49.6 &   7 h 54 m & $32^{\circ}$ $36^{\prime}$\\
2 & 136  & 170.2 & 16.6 & 2455256.64166 & 226.3 & 73.4 &   13 h 25 m & $33^{\circ}$ $49^{\prime}$\\
3 & 276 & 192.9 & 26.0 & 2455708.26792 & 249.0 & 64.0 &   9 h 08 m & $32^{\circ}$ $45^{\prime}$\\
4 & 225 & 235.7 & 23.5 & 2456044.54870 & 291.8 & 66.5 &   13 h 35 m & $49^{\circ}$ $57^{\prime}$\\
5 & 136 & 264.8 & 2.6 & 2456052.38119 & 320.9 & 87.4 &   12 h 14 m& $48^{\circ}$ $18^{\prime}$\\
\hline
\end{tabular}
\end{center}
\end{table}
\begin{figure}[h]
\begin{center}
\includegraphics[width=\columnwidth]{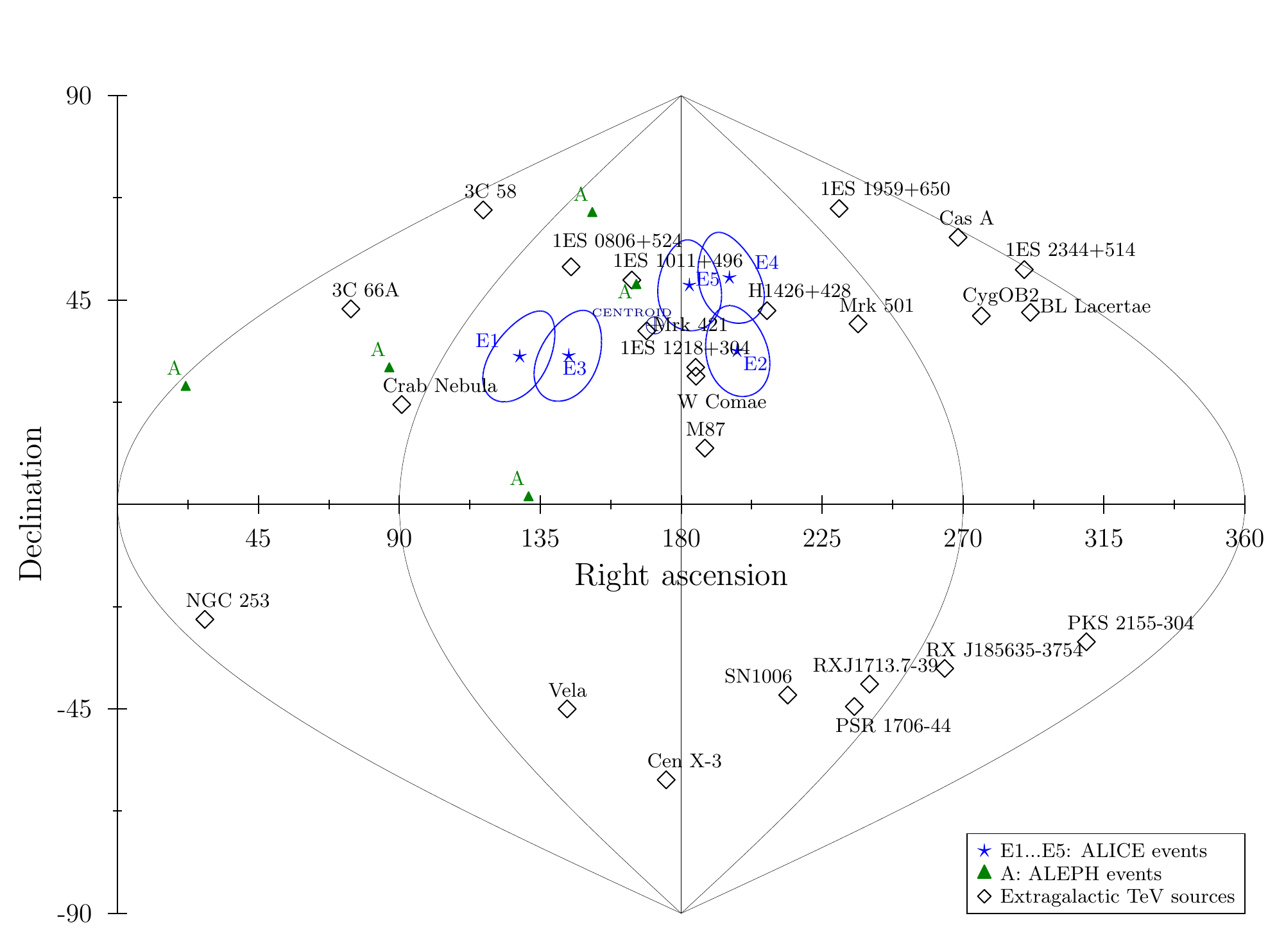}
\vspace{-5mm}
\caption{Five high-multiplicity ALICE muon events in the equatorial reference frame ($\alpha$, $\delta$). For illuctrative purposes we have included also the locations of the five ALEPH events presented in~\cite{Avati:2000mn}. Most known extragalactic TeV sources in the sky (blazars, SNRs, radio galaxies) are also shown (cf.,~\cite{Horan:2004},~\cite{Turley:2016}; note that the Mrk 421 blazar is the source located very close to the centroid of the five considered events.\label{fig1}}
\end{center}
\end{figure}
\begin{figure}[h]
\vspace{0.7cm}
\begin{center}
\includegraphics[width=\columnwidth]{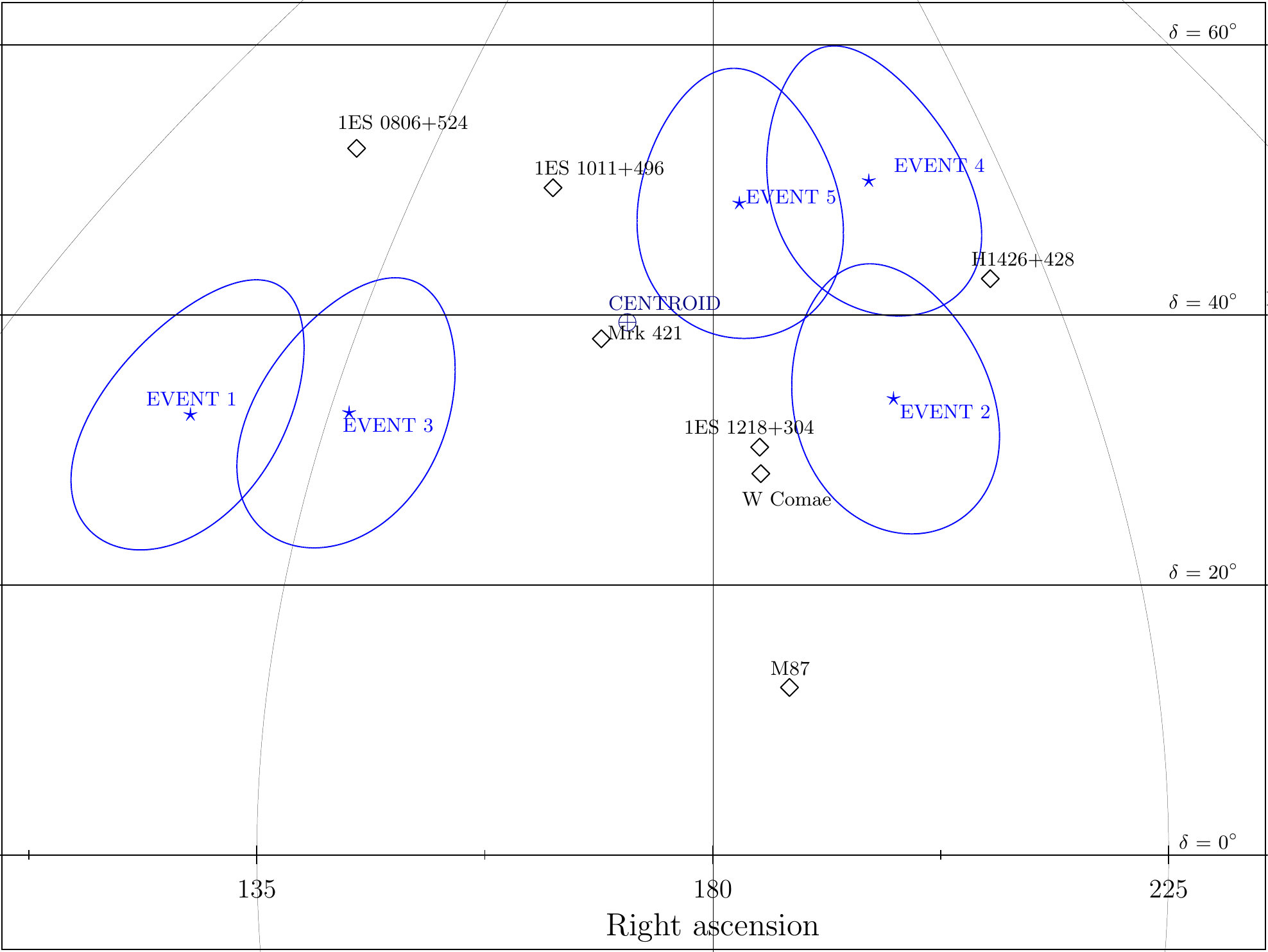}
\caption{Enlarged part of Figure~\ref{fig1} showing in more detail the five high-multiplicity ALICE muon events in the equatorial reference frame ($\alpha$, $\delta$).\label{fig1a}}
\end{center}
\end{figure}
\begin{figure}[h]
\includegraphics[width=\columnwidth]{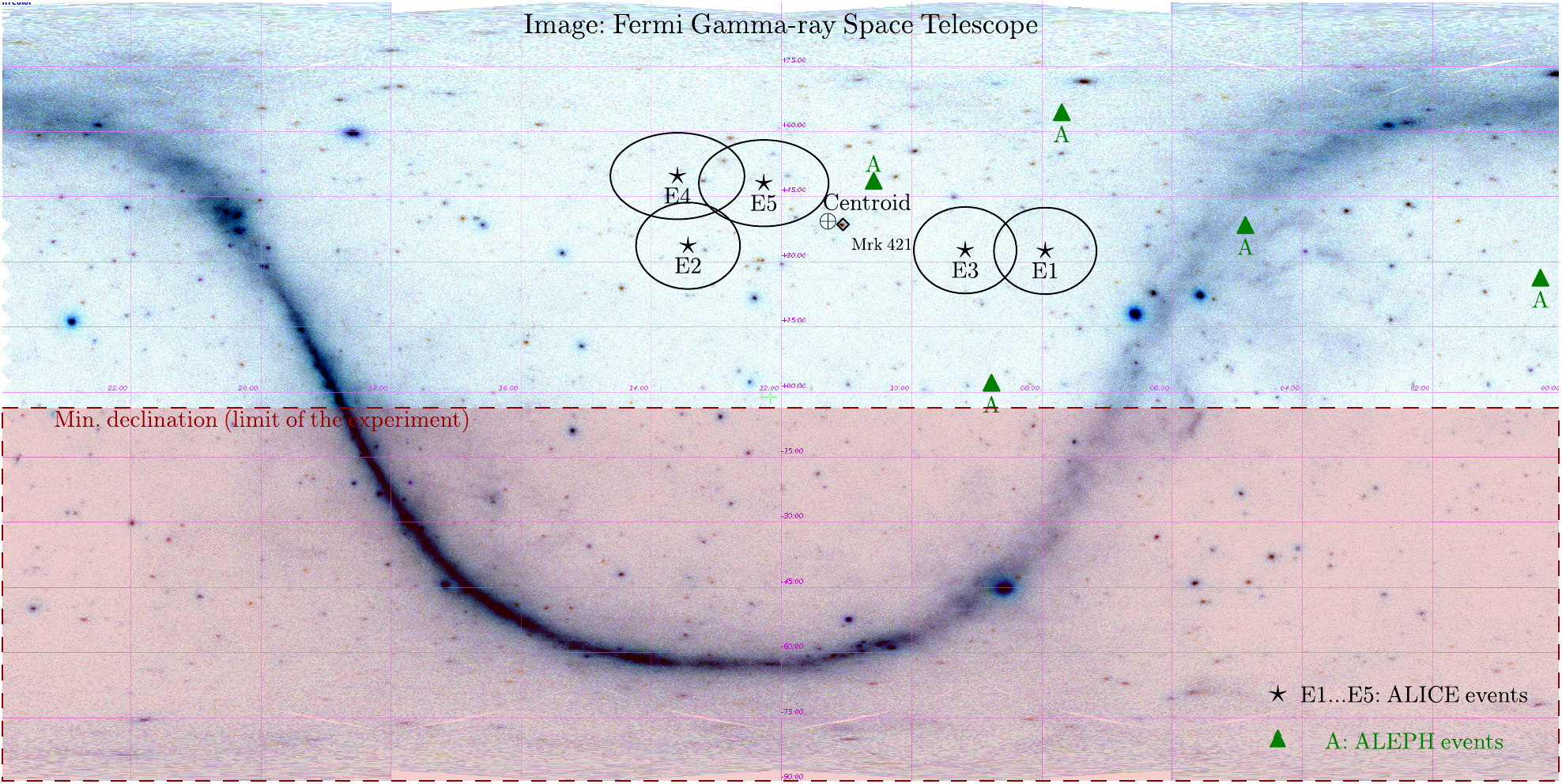}
\caption{Five high-multiplicity ALICE muon events and five old ALEPH events (\cite{Avati:2000mn}) are shown together. All ALICE events and four ALEPH events are located close to the galactic pole (far from the galactic plane); one ALEPH event is located near to the galactic plane. Background: Inverted (negative) image of the Fermi telescope mosaic. The minimum declination limit (due to the restricted zenith angle in the experiment) is marked by a horizontal line. The area in the southern sky not covered by the experiment is marked by a rectangle (filled).\label{fig2}}
\end{figure}

Taking into account the orientation of the axes, these coordinates can be transformed into a horizontal reference frame: $(\Phi,\theta)\to(Az,h)$, where $Az$ is the Azimuth used in navigation/astronomy (in this particular case: $Az=\Phi+\alpha_{0}$), and  $h$ is the elevation above the horizon (where $h=90^{\circ}-\theta$). Finally, we can transform the coordinates from the horizontal to the equatorial reference frame: $(Az,h)\to(\alpha,\delta)$. Horizontal reference frames are related to the observer position and are not ideal choice for use in our issue. For this reason, the equatorial system of coordinates ($\alpha$, $\delta$) is more convenient to specify the direction of our potential sources of events.
This means that we can deduce, for the first time in this kind of experiment, the directionality of the observed muon bundles, which could presumably tell us their possible source. For this purpose we use $5$ bundles of the highest muon multiplicities.

Using the usual spherical triangle formulas or rotation matrices one can convert between the horizontal and equatorial coordinates. To determine the orientation of the equatorial frame some additional parameters are required, like the geographical location of the detector ( $\lambda=6.0208^{\circ}E$, $\phi=46.2516^{\circ}N$) and the universal time (UTC) or local sidereal time ($T^{\star}$) of each observation/event. In this respect it is important to note that each set of $(\alpha,\delta)$ is related to the individual epoch of observation. Because the obtained coordinates relate to individual epochs of observation, in the end we converted all values to J2000.0: $\left( \alpha_{epoch}, \delta_{epoch} \right) \to \left( \alpha_{2000}, \delta_{2000}\right)$. Taking into account the assumed level of accuracy, these corrections are small, but they unify the reference system used for comparison with the coordinates of potential sources. The above information is summarized in Table \ref{Table1}.

To summarize: For the five high-multiplicity ALICE muon events we obtained the celestial equatorial coordinates $(\alpha_{2000},\delta_{2000})$ and the estimated limits of their positional errors are $\sim 10^{\circ}$. They are
shown as circles in Figures \ref{fig1}, \ref{fig1a} and \ref{fig2}. From the dispersion of the distribution of each multi-muon event and the possible errors in the estimation of the $Az$, $h$ and $\alpha_{0}$ angles we assumed that the approximate limit of the positional error is $\sim 10^{\circ}$ for each event. These error limits are shown in Figures \ref{fig1}, \ref{fig1a} and \ref{fig2} as circles. The centroid (average location) of the five considered events is marked by a crossed circle in Figures \ref{fig1} and \ref{fig2}. The coordinates of this centroid are: $\alpha_{centr} = 11~h~15~m$ , $\delta_{centr} = 39^{\circ}$ $29^{\prime}$.

We can conclude therefore that the distribution of the directions of these five ALICE events suggests their possible extragalactic provenance. Most of them are located in the vicinity of the galactic north pole, far from the galactic plane. If we temporarily assume that all these events have the same source, the most probable source would be the blazar Mrk 421 which is located close to the centroid of the five events. It is one of the brightest blazars known, with major outbursts, active at all wavebands \cite{Horan:2004}. However, there are also many other known extragalactic high energy sources located close to the coordinates of our events \cite{Horan:2004}, \cite{Turley:2016} and, at the moment, they cannot be excluded as potential sources.

However, there are additionally five detected some time ago high multiplicity events from the ALEPH experiment (\cite{Avati:2000mn}), which should be also accounted for. Unfortunately, in this case only partial information about the coordinates of particular events was published (like muon density, zenith angle and primary energy, all are listed in \cite{Avati:2000mn}, cf., Table 2 there). Therefore, due to the absence of all necessary information, the more detailed comparison of all events is not possible. On the other hand, ALEPH events are presented on the map in the galactic reference frame (cf., \cite{Avati:2000mn}, right part of Figure 12) what allows to retrieve their approximate galactic coordinates. We transformed therefore these coordinates from the galactic to the equatorial reference frame: $\left(l, b\right) \to \left( \alpha_{2000}, \delta_{2000}\right)$. The approximate location of these five ALEPH events is then shown in Figures \ref{fig1} and \ref{fig2} (marked by triangles). Note that the dispersion of ALEPH events appear to be greater than dispersion of ALICE events. Note also that although four of them are located close to the ALICE events, there remains one event located near the galactic plane. This observation weakness therefore our previous conclusion concerning the anisotropy of the observed events; more further experimental results are obviously needed.

\section{Summary and conclusions}
\label{sec:conc}

There is an ongoing discussion concerning the possibility of the existence in the Universe of some stable forms of strange quark matter and the feasibility of its detection. Whereas until now we do not have fully convincing observations of SQM,  nevertheless there are numerous (and still growing) indications of the possible existence of strange stars, or even of the existence of quark stars. Strangelets could originate, for example, from violent collisions between such objects.
The recent results of ALICE reporting the existence of bundles of high-multiplicity muons could serve, notwithstanding the weakness of this signal (and still existing controversy of its source, as was mentioned before), as a direct (almost) signal of stable lumps of SQM called strangelets arriving at the Earth from outer space.
This is how we have perceived it, based on our previous experience with the search for strangelets in muonic data from the previous CosmoLEP project~\cite{Ryb,Ryb:2002mww,Ryb:2004rr,Ryb:2006mww}, and this is the main rationale behind our work. As a result, we have found that whereas the low multiplicities of muon groups measured by the CERN ALICE experiment favour light nuclei as primaries and the medium multiplicities show behaviour specific for heavier primaries, the muon groups (bundles) of really high multiplicities (of the order of ~$\sim 100$) apparently cannot be described by the common interaction models. We have shown here that the situation can be rescued by allowing for a relatively small (of the order of $10^{-5}$ of the total primary flux) admixture of strangelets of the same total energy. Our estimation of their flux does not contradict the results obtained recently by the SLIM Collaboration~\cite{Sahnoun:2008mr}.  Finally, we would like to stress here that the specific features of the ALICE data allowed us to estimate, for the first time, also the directionality of the five events of the highest muon multiplicities, and to identify their most probable extragallactic source(s). In this respect the previous results provided by ALEPH CosmoLep experiment are not so exact. Nevertheless, they seem to suggest similar conclusion but in much weakened form. To reach more firm conclusion one still needs more data on high multiplicity muonic bundles. We hope that our investigation will provide a new impact for further investigations of this problem\footnote{At this point we would like to bring attention of the reader to the most recent review~\cite{Deligny:2017} summarizing the searches for the anisotropies and the sources of highest energy cosmic rays which have been carried out by the Pierre Auger and Telescope Array collaborations during the past decade. Their data show region with the high concentration of events (the hot spot) in the Northern hemisphere.  We have checked that our centroid of the ALICE muonic events from our Figure~\ref{fig1} situates itself roughly in the hot spot region. Such a correlation would validate the prospects to study astrophysical sources.}. 

\acknowledgments

We are grateful to Dr Katherin Shtejer Diaz and Dr Bruno Alessandro for very useful comments and suggestions concerning estimates of celestial coordinates of high-multiplicity muon events and for providing us with data on their timestamps. We would also like to thank warmly Dr Nicholas Keeley for reading the manuscript.


\end{document}